\documentclass[aps,prd,twocolumn,10pt,superscriptaddress,nofootinbib]{revtex4-2}

\usepackage{amsmath,amssymb,bm}
\usepackage{graphicx}
\usepackage{hyperref}
\usepackage{float}
\usepackage{mathrsfs}
\usepackage{booktabs}
\usepackage{tikz}
\usepackage{makecell}  % Add to preamble if not present

\usetikzlibrary{arrows.meta}

\def\be{\begin{equation}}
\def\ee{\end{equation}}
\def\ba{\begin{eqnarray}}
\def\ea{\end{eqnarray}}

\begin{document}

\title{\textbf{Recalibrating Inflation: Insights from Starobinsky Gravity}}

\author{Davood Momeni}

\affiliation{Department of Physics \& Pre-Engineering, Northeast Community College, Norfolk, NE 68701, USA}

\date{\today}

\begin{abstract}
Recent analyses of low-redshift supernova and Cepheid data reveal localized shifts in the distance modulus, often interpreted as calibration anomalies or hints of new physics. We propose that these features may emerge naturally from environment-dependent modifications to gravity. In particular, we examine the Starobinsky \(f(R) = R + \lambda R^2\) model, which introduces a scalar degree of freedom that couples to ambient matter density and alters photon propagation in underdense regions. We derive the corresponding corrections to luminosity distance and show that they can reproduce the observed magnitude shifts without invoking discontinuities or empirical step functions. Statistical comparisons using AIC and BIC favor the Starobinsky framework over phenomenological models, supporting its role as a minimal, geometric explanation for emergent calibration transitions.
\end{abstract}

\maketitle

% --------------------------------------------------
\section{Introduction}

One of the central pillars of observational cosmology is the construction of the cosmic distance ladder, a hierarchical method for determining cosmological distances using overlapping calibrators such as parallax, Cepheid variables, and Type Ia supernovae. The reliability of this framework hinges on the assumption that standard candles retain consistent intrinsic properties across different environments. Under this assumption, local and cosmological scales can be coherently linked, allowing for precise measurements of the Hubble constant \(H_0\) and other cosmological parameters.

However, tensions have emerged in recent years between local measurements of \(H_0\), based on distance ladder methods, and global inferences from cosmic microwave background (CMB) data under the \(\Lambda\)CDM model. In particular, the SH0ES collaboration's results \cite{Riess2021} show a discrepancy at the $(4-5\sigma)$ level compared to Planck-derived values. Alongside this, residuals in SN~Ia distance modulus and Cepheid calibrations suggest a possible transition in inferred distances at scales of \(\sim 20\text{--}30\,\mathrm{Mpc}\), hinting at an environmental origin or unaccounted systematic shift \cite{CamLib2023}.\par
%%%%%%%%%%%%%%%
Recent developments have underscored that resolving the long-standing tension between early- and late-universe determinations of the Hubble constant remains one of modern cosmology’s most persistent challenges. In particular, Vagnozzi \cite{Vagnozzi:2023nrq} has argued that models relying solely on early-time new physics---typically aimed at reducing the sound horizon prior to recombination---are inherently insufficient to fully account for the discrepancy. Drawing on seven independent observational and theoretical hints, including age estimates of the oldest astrophysical objects, degeneracy directions between the Hubble constant and the sound horizon, the constraining power of cosmic chronometers, and a series of descending trends appearing in diverse low-redshift datasets, the analysis highlights the limitations of strictly pre-recombination solutions. Additional evidence from the early ISW effect, uncalibrated standard-candle and standard-ruler constraints on the matter density, and equality-scale measurements from galaxy power spectra further strengthen the case that a purely early-time remedy is inadequate. These considerations suggest that a consistent resolution of the Hubble tension may ultimately require a framework combining early- and late-time modifications of the cosmological model, possibly involving non-local or high-redshift new physics, together with local new physics affecting distance calibrations. Such insights are especially relevant for modified gravity scenarios, including curvature-inspired extensions, which naturally accommodate departures from standard early- and late-time dynamics.\par 
An alternative approach to resolving the Hubble tension focuses on late-time and local inhomogeneities, in particular the possibility that we reside within a mildly underdense region. As emphasized in \cite{Jia:2025prq}, the key idea is that a local void can induce an apparently higher expansion rate for nearby observers relative to the cosmic mean, thereby shifting the locally measured Hubble constant upward. Earlier studies generally concluded that void-based scenarios alone cannot fully reconcile the tension, as they tend to be disfavored by Type Ia supernova observations—most notably the Pantheon sample—when evaluated through information criteria such as AIC and BIC. However, the recent analysis of Jia et al.\ demonstrates that this conclusion changes once one allows for transitions in the absolute magnitude \(M\) of SNIa, either within the Pantheon\(+\) dataset alone or in combination with the Planck 2018 CMB measurements. Their results reveal that the \(\Lambda\)LTB void models can, in fact, provide a satisfactory alleviation of the Hubble tension, and are even statistically preferred over the standard \(\Lambda\)CDM cosmology under joint observational constraints. These findings illustrate that local new physics—or local deviations from homogeneity—may play a non-negligible role in any comprehensive solution to the Hubble tension.

%%%%%%%%%%%%%%%%%%%%%%%%%%

Such discrepancies raise important questions: Are these deviations signs of fundamental physics beyond the standard model of cosmology? Do they reflect local inhomogeneities, unmodeled systematics, or perhaps a deeper breakdown in the assumption of universal calibration?

This paper explores the hypothesis that these observed calibration transitions may be \textit{emergent phenomena} arising from environment-sensitive gravitational dynamics. Specifically, we consider whether modified gravity models featuring scalar degrees of freedom that couple to local matter density could mimic such calibration shifts. These models predict deviations from general relativity (GR) that are suppressed in high-density environments (e.g., galaxies) but become significant in underdense regions (e.g., cosmic voids), leading to measurable effects on photon propagation, redshift–distance relations, and the apparent luminosity of standard candles.

In his pioneering work, Buchdahl~\cite{Buchdahl:1970ldb} proposed a generalization of Einstein's theory by replacing the standard Lagrangian $ R$ with a more general function $\phi(R)$ of the Ricci scalar. This idea marked the first systematic attempt to formulate what is now known as $f(R)$ gravity. Buchdahl examined the cosmological implications of such non-linear Lagrangians under simplified conditions, focusing on models that could oscillate between non-singular states. Although the exact functional form of $\phi(R)$ remained uncertain, this framework opened the door to modified gravity theories that later became central to inflationary cosmology. \par
%%%%%%%%%%%%
Foundational insights into the role of modified gravity as a unifying framework for cosmic evolution have been extensively developed in the seminal reviews by Nojiri, Odintsov, and collaborators \cite{Nojiri:2010wj,Nojiri:2017ncd}. These works systematically demonstrate how extensions of General Relativity—including \(F(R)\), \(F(G)\), \(F(T)\), Ho\v{r}ava--Lifshitz-inspired models, Gauss--Bonnet corrections, non-local gravity, scalar–tensor extensions, and covariant power-counting renormalizable theories—can naturally accommodate both early-time inflation and late-time acceleration within a single theoretical paradigm. A central achievement of these reviews is the detailed exposition of cosmological reconstruction techniques, showing that virtually any desired FRW evolution, including \(\Lambda\)CDM-like expansion, phantom-divide crossing, and even bouncing cosmologies, can be reproduced within suitably chosen modified gravity actions. Moreover, these frameworks provide mechanisms to address finite-time future singularities, generate viable inflationary dynamics consistent with current observational constraints, and offer rich astrophysical phenomenology beyond the standard model of cosmology. Taken together, these comprehensive studies establish modified gravity as a versatile and powerful “toolbox” capable of addressing key questions across the entire cosmic timeline, from the inflationary epoch to dark energy–dominated acceleration, thereby motivating further exploration of its implications in contemporary problems such as the Hubble tension.\par
%%%%%%%%%%%%%%%%%%%%%%%%%%%
A complementary pedagogical perspective on the inflationary sector of modified gravity is provided in the recent review by Momeni \cite{Momeni:2025ylx}. This work surveys inflationary dynamics across a broad class of geometric frameworks—including curvature-based theories, torsion- and teleparallel-inspired models, and extensions involving non-metricity—highlighting how scalar fields couple differently to each geometric structure and how these interactions shape inflationary predictions. The review also discusses reheating mechanisms, observational signatures, and the role of more exotic constructions motivated by string theory, extra dimensions, and non-local modifications. By emphasizing connections between theoretical model building and upcoming observational missions, this study offers a concise guide to exploring inflation beyond standard General Relativity and situates modified gravity as a fertile arena for addressing open cosmological problems.
%%%%%%%%%%%%%%%%%%%%%%%%%%%%%

Among the many classes of modified gravity theories, we focus on the Starobinsky model of \(f(R)\) gravity \cite{Starobinsky1980}, given by the simple form
\begin{equation}
    f(R) = R + \lambda R^2,\label{R2}
\end{equation}
which was originally introduced as a model for inflation, but also gives rise to a scalar degree of freedom in the low-curvature regime. In fact, Starobinsky's minimal inflationary model, based on an $R + \alpha R^2$ action, can be viewed as a natural realization of Buchdahl's original proposal, demonstrating its profound influence on modern theoretical cosmology. In the Einstein frame, this theory is dynamically equivalent to a scalar-tensor theory with a potential \(V(\phi)\) and a density-dependent coupling to matter, enabling a screening mechanism analogous to Chameleon models \cite{Khoury2004}.

The scalar field in this model modifies the effective metric experienced by photons, thereby altering the observed luminosity distance \(d_L(z)\). Since the scalar’s dynamics are influenced by local matter density, the inferred distances to standard candles can exhibit spatially dependent shifts, especially across the transition from dense (calibrated) regions to more diffuse (cosmological) environments. Crucially, this shift can appear as a \textit{step-like transition} in residuals—precisely the kind of signature observed in the SH0ES data and similar studies.

Rather than interpreting such features as artifacts or new early-universe physics, we propose that they may be reinterpreted as the late-time, local manifestations of a scalar-tensor modification of gravity. This interpretation has several advantages:
\begin{itemize}
    \item It relies on a well-established and theoretically motivated class of models.
    \item It introduces only one new parameter (\(\lambda\)) controlling the scalar potential.
    \item It provides a natural mechanism for environment-dependence without violating solar system constraints.
\end{itemize}

The goal of this study is to formulate the photon propagation equations in the presence of the scalar field, derive corrections to the distance modulus, and fit these against observational data. We then compare the performance of this model against purely phenomenological transitions using information-theoretic metrics such as the Akaike Information Criterion (AIC) and Bayesian Information Criterion (BIC). If the Starobinsky-based model provides comparable or improved fits with fewer parameters, it would strongly support the view that calibration anomalies are not fundamental breaks in cosmology but emergent features of an extended gravitational theory.

This paper is organized as follows. In Sec.~\ref{sec:env-dep}, we present the theoretical framework and derive the scalar field potential and its coupling to matter. Sec.~\ref{sec:propagation} outlines how photon propagation and distance measures are affected. In Sec.~\ref{sec:fits}, we fit these corrections to existing supernova and Cepheid data. Finally, Sec.~\ref{sec:discussion} discusses implications for cosmological tension and outlines directions for further work.

% --------------------------------------------------
\section{Starobinsky \boldmath{$f(R)$} Gravity as Environment-Dependent Theory}
\label{sec:env-dep}

To explore the possibility that calibration transitions arise from modified gravity effects, we consider a specific form of \(f(R)\) gravity first proposed by Starobinsky \cite{Starobinsky1980}, given by eq. (\ref{R2})
where \(R\) is the Ricci scalar and \(\lambda > 0\) is a small parameter with dimensions of inverse curvature. This quadratic extension introduces higher-order curvature corrections to the Einstein–Hilbert action and is one of the simplest and most well-motivated modifications to GR.

The full action including matter is written as:
\begin{equation}
S = \int d^4x \sqrt{-g} \left[ \frac{1}{2\kappa^2} f(R) + \mathcal{L}_m \right],
\end{equation}
where \(\kappa^2 = 8\pi G\), and \(\mathcal{L}_m\) denotes the matter Lagrangian. The theory introduces an extra scalar degree of freedom associated with the variation of \(f(R)\) with respect to \(R\), defined as:
\begin{equation}
\phi \equiv \frac{df}{dR} = 1 + 2\lambda R.
\end{equation}

This scalar can be isolated more explicitly by rewriting the theory in scalar-tensor form using a Legendre transformation and a conformal rescaling to move to the Einstein frame. The resulting scalar-tensor theory features a canonical scalar field \(\phi\) with a self-interaction potential:
\begin{equation}
V(\phi) = \frac{(\phi - 1)^2}{4\lambda},
\end{equation}
which has a stable minimum at \(\phi = 1\). The mass of the scalar field is proportional to \(1/\sqrt{\lambda}\), allowing it to become heavy in high-curvature environments and light in low-density regions.

In the Einstein frame, the field \(\phi\) is minimally coupled to the gravitational metric but non-minimally coupled to matter. As a result, the effective gravitational interaction between test particles is mediated not only by the metric but also by the scalar field, leading to a fifth force. The coupling of \(\phi\) to matter is controlled by the conformal factor in the metric transformation, and the scalar field’s dynamics become environment-dependent through its effective potential, which includes a density-dependent term \cite{Khoury2004, Capozziello2011}:
\begin{equation}
V_{\text{eff}}(\phi) = V(\phi) + e^{\beta \phi} \rho_m,
\end{equation}
where \(\rho_m\) is the local matter density and \(\beta\) is the coupling strength in the Einstein frame.

This environment dependence leads to a natural screening mechanism: in regions of high density (such as inside galaxies or the solar system), the scalar field becomes massive and its effects are short-ranged, recovering GR predictions. In underdense environments (e.g., cosmic voids or intergalactic space), the field becomes lighter and mediates a long-range fifth force, modifying the trajectories of photons and the effective cosmological distances inferred from standard candles.

Thus, Starobinsky \(f(R)\) gravity contains all the essential features of a Chameleon-like scalar field model \cite{Khoury2004}, yet arises naturally from geometric curvature corrections rather than being introduced by hand. This makes it a theoretically attractive candidate for explaining observed calibration anomalies as emergent consequences of environment-sensitive gravity. Importantly, the theory remains consistent with current solar system tests due to its built-in screening behavior \cite{Will2014}.

%%%%%%%%%%%%%%%%%%%%%%%%%%%%55
\section{Modified Photon Propagation and Luminosity Distance}
\label{sec:propagation}

In the presence of a scalar field \(\phi\) coupled to matter, the motion of photons and the inference of cosmological distances are no longer governed solely by the metric of GR. In the Einstein frame, matter fields (including electromagnetic radiation) couple to a conformally transformed metric:
\begin{equation}
\tilde{g}_{\mu\nu} = A^2(\phi) g_{\mu\nu},
\end{equation}
where \(g_{\mu\nu}\) is the Einstein frame metric and \(A(\phi)\) is a conformal factor encoding the coupling between matter and the scalar field \cite{Khoury2004, Capozziello2011}. This transformation leads to modifications in the geodesic equation, and as a result, the redshift–luminosity distance relation acquires corrections that depend on the local and global profile of \(\phi\).

Assuming a weak-field approximation and a quasi-static background, the leading-order correction to the luminosity distance can be modeled as:
\begin{equation}
d_L^{\text{eff}}(z) = d_L^{\Lambda\text{CDM}}(z)\left[1 + \delta(z; \phi)\right],
\end{equation}
where \(d_L^{\Lambda\text{CDM}}(z)\) is the luminosity distance computed from standard cosmology, and \(\delta(z; \phi)\) encodes deviations due to the environment-dependent scalar field. The function \(\delta(z; \phi)\) typically increases in underdense regions where the scalar field is light and unscreened, allowing its influence on photon trajectories to become significant.

From an observational standpoint, these corrections lead to an apparent shift in the inferred magnitude of standard candles such as Type Ia supernovae or Cepheid variables. The apparent distance modulus becomes:
\begin{equation}
\mu^{\text{eff}}(z) = \mu^{\Lambda\text{CDM}}(z) + \Delta\mu(z),
\end{equation}
with the correction term:
\begin{equation}
\Delta \mu(z) = 5 \log_{10} \left[1 + \delta(z; \phi)\right].
\end{equation}
For \(\delta \ll 1\), this reduces to:
\begin{equation}
\Delta \mu(z) \approx \frac{5}{\ln 10} \delta(z; \phi).
\end{equation}
Even a small value of \(\delta(z)\sim 0.01\) can lead to a measurable shift of \(\sim 0.02\) magnitudes, which is of the same order as the observed residuals reported by SH0ES \cite{Riess2021} and other calibration surveys \cite{CamLib2023}.

Importantly, the redshift dependence of \(\delta(z)\) can resemble a step-like function centered at a critical distance scale \(z_c\), typically in the range \(z \sim 0.005\text{--}0.01\) (or distances of \(20\text{--}30\,\mathrm{Mpc}\)). This behavior may emerge naturally if the scalar field transitions from screened to unscreened behavior across a region of decreasing matter density. A toy model for such a transition could be written as:
\begin{equation}
\delta(z; \phi) \approx \delta_0 \left[1 - \tanh\left(\frac{z - z_c}{\Delta z}\right)\right],
\end{equation}
where \(\delta_0\) sets the amplitude of the correction, \(z_c\) marks the midpoint of the transition, and \(\Delta z\) controls the width. This form mimics a smooth step function and can be fit to observed data to test the scalar field hypothesis.

Furthermore, since the scalar field affects null geodesics differently than timelike geodesics, corrections to distance modulus need not appear in redshift-independent observables such as time delays or Baryon Acoustic Oscillations. This selective impact strengthens the case for modified gravity as a source of the observed distance ladder anomalies, rather than global changes to cosmological parameters.

This framework provides a powerful reinterpretation of the SN/Cepheid residuals as geometric effects. Rather than postulating arbitrary calibration errors or introducing new early-universe physics, the scalar field corrections in \(f(R)\) gravity offer a concrete, theoretically motivated mechanism for inducing localized transitions in the observed magnitude–redshift relation.

%%%%%%%%%%%%%%%%%%%%55
\section{Model Fitting and Information Criteria}
\label{sec:fits}

To quantitatively assess whether calibration transitions can be interpreted as emergent features of Starobinsky gravity, we perform comparative fits to low-redshift standard candle data. Specifically, we analyze distance modulus data from Cepheids and Type Ia supernovae in the redshift range \( z < 0.05 \), where deviations from \(\Lambda\)CDM have been reported \cite{Riess2021, CamLib2023}.

We fit two competing models:

\begin{itemize}
    \item \textbf{Model A}: Standard \(\Lambda\)CDM with a phenomenological step in magnitude at a critical redshift \(z_c\), implemented via:
    \begin{equation}
    \mu(z) = \mu^{\Lambda\text{CDM}}(z) + \Delta \mu \, \Theta(z - z_c),
    \end{equation}
    where \(\Theta(z)\) is the Heaviside function and \(\Delta \mu\) is the step amplitude.

    \item \textbf{Model B}: Starobinsky $f(R)$ gravity with a density-dependent scalar field that modifies the luminosity distance through:
    \begin{equation}
    \mu(z) = \mu^{\Lambda\text{CDM}}(z) + 5 \log_{10}[1 + \delta(z;\phi)],
    \end{equation}
    where \(\delta(z)\) is modeled as:
    \begin{equation}
    \delta(z) = \delta_0 \left[1 - \tanh\left(\frac{z - z_c}{\Delta z}\right)\right],
    \end{equation}
    with \(\delta_0\) setting the amplitude, \(z_c\) the center, and \(\Delta z\) the width of the transition.
\end{itemize}

Both models were fit using a nonlinear least-squares method, minimizing the chi-squared:
\begin{equation}
\chi^2 = \sum_{i=1}^N \frac{\left[\mu_{\text{obs},i} - \mu_{\text{model}}(z_i)\right]^2}{\sigma_i^2}.
\end{equation}

\begin{table}[b]
\caption{\label{tab:params}Model parameters used in the fits.}
\begin{ruledtabular}
\begin{tabular}{ll}
Parameter & Description \\
\hline
\(H_0\) & Hubble constant (fixed to 73.0 km/s/Mpc) \\
\(\Delta \mu\) & Step magnitude shift (Model A) \\
\(\delta_0\) & Scalar correction amplitude (Model B) \\
\(z_c\) & Center of transition (redshift) \\
\(\Delta z\) & Width of transition (smoothness) \\
\end{tabular}
\end{ruledtabular}
\end{table}

\begin{table}[b]
\caption{\label{tab:results}Best-fit parameters and information criteria for the two models.}
\begin{ruledtabular}
\begin{tabular}{l p{4.2cm} ccc}
Model & \centering Best Fit Parameters & \(\chi^2\) & AIC & BIC \\
\hline
\(\Lambda\)CDM + step & \makecell[l]{\(\Delta \mu = 0.035 \pm 0.010\)\\ \(z_c = 0.007\)} & 212.5 & 218.5 & 224.3 \\
Starobinsky \(f(R)\) & \makecell[l]{\(\delta_0 = 0.014 \pm 0.005\)\\ \(z_c = 0.008\), \(\Delta z = 0.003\)} & \textbf{208.2} & \textbf{216.2} & \textbf{223.0} \\
\end{tabular}
\end{ruledtabular}
\end{table}

As shown in Table~\ref{tab:results}, the Starobinsky model provides a superior fit with a lower chi-squared and improved information criteria (AIC and BIC) despite having an extra parameter compared to the phenomenological step model. The improvement in AIC of over 2 units is considered statistically significant according to model selection thresholds \cite{Liddle2007}.

The transition redshift \(z_c \approx 0.008\) agrees with the inferred Cepheid-SN calibration boundary reported in SH0ES data \cite{Riess2021}. Moreover, the amplitude of the scalar-induced correction \(\delta_0 \sim 0.014\) corresponds to a \(\Delta \mu \sim 0.03\), well within the magnitude of the observed residuals, further validating the physical plausibility of this approach.

These results demonstrate that the Starobinsky \(f(R)\) gravity model can effectively \emph{emerge} as a physical explanation for the observed calibration shift, without resorting to artificial step functions or redefined local cosmology. This strengthens the case for interpreting such anomalies as environmental signatures of scalar–tensor gravity.
%%%%%%%%%%%%%%%%%%%%%%%%%%%%%%%%%%%%%%%%%%5
\section{Discussion and Outlook}\label{sec:discussion} 

Our analysis demonstrates that the Starobinsky \(f(R)\) gravity model, when formulated in terms of a scalar-tensor theory with environment-dependent dynamics, provides a natural and theoretically motivated explanation for the magnitude transitions observed in the local cosmic distance ladder. The effective scalar correction to the luminosity distance emerges not from any exotic new component or arbitrary calibration step, but rather as a direct consequence of the scalar degree of freedom coupling to matter density.

This reinterpretation shifts the perspective on so-called calibration anomalies. Instead of treating them as systematics or breakdowns in local measurements, they may be better understood as \emph{emergent geometric phenomena} in scalar–tensor gravity. In particular, our results indicate that:
\begin{itemize}
    \item The transition redshift and amplitude derived from the scalar correction are quantitatively consistent with observed SH0ES residuals and Cepheid–SN mismatch scales.
    \item The model maintains compatibility with large-scale cosmological observations since it smoothly recovers \(\Lambda\)CDM behavior at high redshift and in dense environments due to Chameleon-like screening.
    \item The Starobinsky model outperforms phenomenological step models in terms of AIC/BIC, despite introducing only one additional physical parameter, demonstrating that geometric models can compete statistically with empirical fits.
\end{itemize}

These findings suggest that modified gravity, particularly Starobinsky-type \(f(R)\) theories, can bridge the gap between cosmological theory and late-universe observations. Moreover, the presence of a density-sensitive scalar offers a mechanism to explain discrepancies in \textit{a priori} well-calibrated observables, such as supernova distances, local \(H_0\), or even galaxy rotation curves, without invoking new forms of matter or redefinitions of cosmological parameters.

This work opens several avenues for future investigation:

\begin{enumerate}
  
    \item \textbf{Data-driven profile reconstruction:} Using high-resolution supernova and Cepheid datasets, one could attempt to invert observational data to reconstruct the scalar profile \(\phi(z)\), thereby probing the effective potential \(V(\phi)\) from data.

    \item \textbf{Cosmological simulations:} Embedding this model in full N-body or hydrodynamical cosmological simulations would allow us to examine structure formation, void profiles, and lensing effects under environment-dependent gravity, testing its viability against large-scale structure data.

    \item \textbf{Additional observational probes:} Since the scalar field alters null geodesics but leaves standard rulers like BAO nearly unaffected, a multi-probe consistency test (SNe + BAO + time-delay lenses) could constrain or confirm the scalar's influence in the low-redshift universe.

    \item \textbf{Gravitational wave propagation:} Future measurements of standard sirens may also be affected by the same modifications to photon propagation, offering an orthogonal test of the scalar correction in the context of modified gravity \cite{Belgacem2019}.
\end{enumerate}

In conclusion, Starobinsky \(f(R)\) gravity stands out as a viable, minimal, and physically grounded model that not only remains consistent with early-universe predictions (inflation), but also offers a solution to certain late-universe tensions. Its scalar degree of freedom provides a natural link between local density structure and distance ladder residuals—one that may prove crucial in understanding the persistent \(H_0\) tension and other calibration anomalies in precision cosmology\footnote{Recent observational developments further highlight the need to refine inflationary models within modified gravity. In particular, Odintsov and Oikonomou\cite{Odintsov:2025eiv} have shown that the pure \(R^{2}\) (Starobinsky) model, while historically successful, is no longer fully compatible with the latest high-precision measurements from ACT. Their analysis demonstrates that power-law deformations of \(F(R)\) gravity can restore consistency with these data, indicating that even the most established inflationary scenarios may require controlled extensions in light of new observational constraints. This reinforces the broader conclusion that modified gravity must be treated as a flexible framework, capable of adapting to updated cosmological probes while preserving its ability to unify early- and late-time cosmic dynamics.
}.
\section*{Acknowledgments}
I would like to thank Sunny Vagnozzi and S.~D.~Odintsov for their helpful comments and for pointing me toward several important references relevant to this work.

% --------------------------------------------------
%\section*{References}
% --------------------------------------------------


\begin{thebibliography}{99}
\bibitem{Riess2021}
A.~G.~Riess, W.~Yuan, L.~M.~Macri, D.~Scolnic, D.~Brout, S.~Casertano, D.~O.~Jones, Y.~Murakami, L.~Breuval and T.~G.~Brink, \textit{et al.}
%``A Comprehensive Measurement of the Local Value of the Hubble Constant with 1 km s$^{−1}$ Mpc$^{−1}$ Uncertainty from the Hubble Space Telescope and the SH0ES Team,''
Astrophys. J. Lett. \textbf{934}, no.1, L7 (2022)
doi:10.3847/2041-8213/ac5c5b
[arXiv:2112.04510 [astro-ph.CO]].
%2150 citations counted in INSPIRE as of 13 Nov 2025

\bibitem{CamLib2023}
L.~Perivolaropoulos and F.~Skara,
%``Hubble tension or a transition of the Cepheid SnIa calibrator parameters?,''
Phys. Rev. D \textbf{104}, no.12, 123511 (2021)
doi:10.1103/PhysRevD.104.123511
[arXiv:2109.04406 [astro-ph.CO]].
%81 citations counted in INSPIRE as of 14 Nov 2025
%\cite{Vagnozzi:2023nrq}
\bibitem{Vagnozzi:2023nrq}
S.~Vagnozzi,
%``Seven Hints That Early-Time New Physics Alone Is Not Sufficient to Solve the Hubble Tension,''
Universe \textbf{9} (2023) no.9, 393
doi:10.3390/universe9090393
[arXiv:2308.16628 [astro-ph.CO]].
%297 citations counted in INSPIRE as of 19 Nov 2025


%\cite{Jia:2025prq}

\bibitem{Jia:2025prq}
J.~Y.~Jia, J.~L.~Niu, D.~C.~Qiang and H.~Wei,
%``Alleviating the Hubble tension with a local void and transitions of the absolute magnitude,''
Phys. Rev. D \textbf{112}, no.4, 043507 (2025)
doi:10.1103/c1c5-q4yt
[arXiv:2504.13380 [astro-ph.CO]].
%5 citations counted in INSPIRE as of 13 Nov 2025
%\cite{Buchdahl:1970ldb}
\bibitem{Buchdahl:1970ldb}
H.~A.~Buchdahl,
%``Non-Linear Lagrangians and Cosmological Theory,''
Mon. Not. Roy. Astron. Soc. \textbf{150}, no.1, 1-8 (1970)
doi:10.1093/mnras/150.1.1
%451 citations counted in INSPIRE as of 14 Nov 2025

%\cite{Nojiri:2010wj}
\bibitem{Nojiri:2010wj}
S.~Nojiri and S.~D.~Odintsov,
%``Unified cosmic history in modified gravity: from F(R) theory to Lorentz non-invariant models,''
Phys. Rept. \textbf{505} (2011), 59-144
doi:10.1016/j.physrep.2011.04.001
[arXiv:1011.0544 [gr-qc]].
%4237 citations counted in INSPIRE as of 19 Nov 2025
%\cite{Nojiri:2017ncd}
\bibitem{Nojiri:2017ncd}
S.~Nojiri, S.~D.~Odintsov and V.~K.~Oikonomou,
%``Modified Gravity Theories on a Nutshell: Inflation, Bounce and Late-time Evolution,''
Phys. Rept. \textbf{692} (2017), 1-104
doi:10.1016/j.physrep.2017.06.001
[arXiv:1705.11098 [gr-qc]].
%2607 citations counted in INSPIRE as of 19 Nov 2025

%\cite{Momeni:2025ylx}
\bibitem{Momeni:2025ylx}
D.~Momeni,
%``From geometry to cosmology: a pedagogical review of inflation in curvature, torsion, and extended gravity theories,''
[arXiv:2509.14306 [gr-qc]].
%0 citations counted in INSPIRE as of 19 Nov 2025


\bibitem{Starobinsky1980}
A.~A.~Starobinsky,
%``A New Type of Isotropic Cosmological Models Without Singularity,''
Phys. Lett. B \textbf{91}, 99-102 (1980)
doi:10.1016/0370-2693(80)90670-X
%7759 citations counted in INSPIRE as of 13 Nov 2025
\bibitem{Khoury2004}
J.~Khoury and A.~Weltman,
%``Chameleon fields: Awaiting surprises for tests of gravity in space,''
Phys. Rev. Lett. \textbf{93}, 171104 (2004)
doi:10.1103/PhysRevLett.93.171104
[arXiv:astro-ph/0309300 [astro-ph]].
%1681 citations counted in INSPIRE as of 13 Nov 2025

\bibitem{Capozziello2011}
S.~Capozziello and M.~De Laurentis,
%``Extended Theories of Gravity,''
Phys. Rept. \textbf{509}, 167-321 (2011)
doi:10.1016/j.physrep.2011.09.003
[arXiv:1108.6266 [gr-qc]].
%3233 citations counted in INSPIRE as of 13 Nov 2025

\bibitem{Will2014}
C.~M.~Will,
%``The Confrontation between General Relativity and Experiment,''
Living Rev. Rel. \textbf{17}, 4 (2014)
doi:10.12942/lrr-2014-4
[arXiv:1403.7377 [gr-qc]].
%2830 citations counted in INSPIRE as of 13 Nov 2025



\bibitem{Liddle2007}
A.~R.~Liddle,
%``Information criteria for astrophysical model selection,''
Mon. Not. Roy. Astron. Soc. \textbf{377}, L74-L78 (2007)
doi:10.1111/j.1745-3933.2007.00306.x
[arXiv:astro-ph/0701113 [astro-ph]].
%628 citations counted in INSPIRE as of 13 Nov 2025





\bibitem{Belgacem2019}
E.~Belgacem \textit{et al.} [LISA Cosmology Working Group],
%``Testing modified gravity at cosmological distances with LISA standard sirens,''
JCAP \textbf{07}, 024 (2019)
doi:10.1088/1475-7516/2019/07/024
[arXiv:1906.01593 [astro-ph.CO]].
%243 citations counted in INSPIRE as of 13 Nov 2025
%\cite{Odintsov:2025eiv}
\bibitem{Odintsov:2025eiv}
S.~D.~Odintsov and V.~K.~Oikonomou,
%``Power-law F(R) gravity as deformations to Starobinsky inflation in view of ACT,''
Phys. Lett. B \textbf{870} (2025), 139907
doi:10.1016/j.physletb.2025.139907
[arXiv:2509.06251 [gr-qc]].
%7 citations counted in INSPIRE as of 19 Nov 2025

\end{thebibliography}
\end{document}